\documentclass[superscriptaddress,aps,twocolumn,preprintnumbers,showpacs,showkeys,nofootinbib,floatfix]{revtex4-1}
\usepackage{amssymb,amsmath,amsfonts,amsthm,graphicx,psfrag}
\usepackage{hyperref}

\newcommand{\beq}{\begin{eqnarray}}
\newcommand{\eeq}{\end{eqnarray}}

\begin{document}
\preprint{}

\title{Observation of deconfinement in a cold dense quark medium}

\author{V.~G.~Bornyakov}
\affiliation{Institute for Theoretical and Experimental Physics NRC "Kurchatov Institute", Moscow, 117218 Russia}
\affiliation{Far Eastern Federal University, School of Biomedicine, 690950 Vladivostok, Russia}
\affiliation{Institute for High Energy Physics NRC "Kurchatov Institute", Protvino, 142281 Russia}

\author{V.~V.~Braguta}
\affiliation{Institute for Theoretical and Experimental Physics NRC "Kurchatov Institute", Moscow, 117218 Russia}
\affiliation{Far Eastern Federal University, School of Biomedicine, 690950 Vladivostok, Russia}
\affiliation{Moscow Institute of Physics and Technology, Institutskii per. 9, Dolgoprudny, Moscow Region, 141700 Russia}
\affiliation{Bogoliubov Laboratory of Theoretical Physics, Joint Institute for Nuclear Research, Dubna, 141980 Russia}

																																																																																																																																																																																																																																																																																																																																																																																																																																																																																																																																																																																																																																																																																																																																																																																\author{E.-M. Ilgenfritz}
\affiliation{Bogoliubov Laboratory of Theoretical Physics, Joint Institute for Nuclear Research, Dubna, 141980 Russia}

\author{A.~Yu.~Kotov}
\email[]{kotov@itep.ru}
\affiliation{Institute for Theoretical and Experimental Physics NRC "Kurchatov Institute", Moscow, 117218 Russia}
\affiliation{Moscow Institute of Physics and Technology, Institutskii per. 9, Dolgoprudny, Moscow Region, 141700 Russia}
\affiliation{Bogoliubov Laboratory of Theoretical Physics, Joint Institute for Nuclear Research, Dubna, 141980 Russia}

\author{A. V. Molochkov}
\affiliation{Far Eastern Federal University, School of Biomedicine, 690950 Vladivostok, Russia}

\author{A.~A.~Nikolaev}
\email[]{nikolaev.aa@dvfu.ru}
\affiliation{Institute for Theoretical and Experimental Physics NRC "Kurchatov Institute", Moscow, 117218 Russia}
\affiliation{Far Eastern Federal University, School of Biomedicine, 690950 Vladivostok, Russia}

\begin{abstract}
In this paper we study the confinement/deconfinement transition in 
lattice $SU(2)$ QCD at finite quark density and zero temperature.
The simulations are performed on an $32^4$ lattice with rooted 
staggered fermions at a lattice spacing $a = 0.044 \mathrm{~fm}$.
This small lattice spacing allowed us to reach very large baryon 
density (up to quark chemical potential $\mu_q > 2000 \mathrm{~MeV}$)
avoiding strong lattice artifacts.
In the region $\mu_q\sim 1000 \mathrm{~MeV}$ we observe for the first 
time the confinement/deconfinement transition which manifests itself 
in rising of the Polyakov loop and vanishing of the string tension 
$\sigma$. After the deconfinement is achieved at $\mu_q > 1000 \mathrm{~MeV}$, 
we observe a monotonous decrease of the spatial string tension $\sigma_s$
which ends up with $\sigma_s$ vanishing at $\mu_q > 2000 \mathrm{~MeV}$.
From this observation we draw the conclusion that the confinement/deconfinement
transition at finite density and zero temperature is quite different from
that at finite temperature and zero density. Our results indicate that in
very dense matter the quark-gluon plasma is in essence a weakly interacting 
gas of quarks and gluons without a magnetic screening mass in the system,
sharply different from a quark-gluon plasma at large temperature.
\end{abstract}



\maketitle

The knowlegde of the properties of QCD at finite baryon density is very
important for understanding cosmology and astrophysics. 
A thorough experimental study of baryon--rich strongly interacting matter 
is planned at future heavy ion collision experiments FAIR and NICA. Today, 
QCD at high energy density and small baryon density is well explored thanks 
to lattice simulations. Unfortunately, lattice simulations cannot be directly 
applied to the study of properties of the theory at sufficiently large baryon 
density because of the sign problem (for a review see, e.g.~\cite{Muroya:2003qs}).
For this reason one has a rather poor knowledge about the QCD phase diagram
in the region of large baryon density.

There are a lot of phenomenological models which predict different interesting
phenomena in this region of the phase diagram. As examples of such phenomena 
one should mention color flavor locking~\cite{Alford:1998mk}, non-uniform 
phases in dense matter~\cite{Kojo:2009ha}. More conventional and general 
features that people believe in are the restoration of chiral
symmetry~\cite{Alford:1997zt} and deconfinement~\footnote{Notice that the 
confinement/deconfinement and the chiral symmetry breaking/restoration 
transitions are not immediately related~\cite{Suganuma:2017syi}, and in 
dense matter they may take place at different densities.} in dense 
QCD~\cite{McLerran:2007qj} etc. It is rather difficult to estimate systematic 
uncertainties of different phenomenological models under discussion. So, it 
is hardly possible to assess if these phenomena are realized in the real world.

In this paper we are going to study the deconfinement aspects of the 
transition (or transitions) in dense quark matter at low temperature. 
Since in our consideration it is assumed that the temperature is much 
smaller than the baryon chemical potential, $T \ll \mu_b$, there is no 
hope that in the nearest future this region of the phase diagram will be 
reached by standard methods used to overcome the sign problem in simulations 
of $SU(3)$ QCD. Instead of considering thee--color QCD, in this paper we are 
going to describe results of lattice simulation of QCD with the $SU(2)$ gauge
group. This theory is free from the sign problem and it can be directly 
studied on the lattice.

$SU(2)$ and $SU(3)$ gauge theories with fundamental quarks have many 
properties in common. In particular, in both theories confinement/deconfinement
and chiral symmetry breaking/restoration transitions take place at a certain 
non--zero temperature. The mechanism of the fermion mass generation in a dense 
medium and even the formula for the fermion mass gap is the same in both
theories~\cite{Son:1998uk}. In addition, a lot of (ratios of) observables 
are almost independent of the number of colors~\cite{Lucini:2012gg}.

However, these two theories have two important differences.
The first one is that the chiral symmetry breaking pattern for the $SU(3)$
gauge theory differs from that in the $SU(2)$ theory~\cite{Kogut:2000ek}.
The second difference is that baryons in the $SU(3)$ theory contain three
quarks and are fermions whereas is the $SU(2)$ theory baryons contain two 
quarks and are bosons.

Notice that both at sufficiently large chemical potential
and at large temperature the chiral symmetry is restored.
One may argue that in these regions the chiral symmetry breaking 
patterns (which differ in both theories) are not important. Moreover, 
in these regions of the phase diagram, where the relevant degrees of 
freedom are quarks and gluons (what happens at large $\mu$ or $T$), 
it should not be important if baryons are composed of two or three quarks.
These arguments make us believe that lattice simulations of the $SU(2)$
gauge theory can be used not only for a qualitative study of dense matter
but it can also give quantitative predictions for dense $SU(3)$ QCD.

There are several papers devoted to study of dense lattice $SU(2)$ QCD
(see papers~\cite{Kogut:2002cm, Cotter:2012mb, Braguta:2016cpw, 
Holicki:2017psk} and references therein). It is clear that -- in order to 
observe deconfinement in dense matter -- one needs to reach sufficiently 
large values of the quark chemical potential without being hampered by 
lattice artifacts. We believe that the largest values of the quark chemical 
potential $\mu_q \sim 800-1000 \mathrm{~MeV}$ have been safely reached in 
our previous paper~\footnote{Note that throughout this paper we express our 
results as functions of the quark chemical potential $\mu_q$. As we study 
the theory with gauge group $SU(2)$, this corresponds to baryon chemical 
potential $\mu_b=2 \mu_q$.}~\cite{Braguta:2016cpw}.
However, no signatures of the deconfinement phase were observed in the 
previous paper.

In the present paper we continue our study of two-color QCD the region of very
large baryon (quark) density. In particular, we carry out lattice simulations 
with rooted staggered fermions which in the continuum corresponds to $N_f=2$ 
quark flavours. In order to observe condensation of scalar diquarks in a finite
simulation volume we introduced a diquark source term into the lattice action, 
which is controlled by the parameter $\lambda$. In the previous 
paper~\cite{Braguta:2016cpw} we have observed that in the region of large 
baryon density our results only weakly depend on the value of the parameter 
$\lambda$. In the present paper we have investigated in detail the dependence 
of the order parameters on $\lambda$ only for few values of chemical potential 
$\mu_q$ and confirmed that the sensitivity is indeed weak. So, in order to 
reduce the time used for the simulations we restrict our consideration to 
the value $\lambda=0.00075$ which is much smaller than the fermion mass 
$am=0.007$ used in the simulations.

Contrary to our previous study we have used now the Symanzik improved gauge 
action. In addition to the bulk of simulations at finite $\mu_q$, also 
simulations for calibration at $\mu_q=0$ have been performed.~\footnote{We 
used the QCD Sommer scale $r_0=0.468(4) \mathrm{~fm}$ ~\cite{Bazavov:2011nk}
to carry out the scale setting.} In this case our string tension at $\mu_q=0$
amounts to $\sqrt{\sigma_0}=476(5) \mathrm{~MeV}$ at $a = 0.044 \mathrm{~fm}$,
whereas in our previous study~\cite{Braguta:2016cpw} the lattice spacing was 
almost three times larger, $a = 0.112 \mathrm{~fm}$. This change allowed us 
to approach the continuum limit much closer and to reach larger baryon 
densities without being hampered by lattice artifacts. In particular, in
the present paper we reach the region of baryon density corresponding to a 
quark chemical potential $\mu_q > 2000 \mathrm{~MeV}$ which is the largest 
value ever reached in lattice simulations of $SU(2)$ QCD.

The simulations are performed on a $32^4$ lattice (compare with the
lattice $16^3\times32$ in our previous study \cite{Braguta:2016cpw}).
Numerical simulations in the region of large baryon density require 
considerable computer resources. For this reason, for the present paper 
we conducted our study at a pion mass of $m_{\pi}=740(40) \mathrm{~MeV}$, 
a value which is larger than that used in \cite{Braguta:2016cpw}.
We will preferentially decrease the pion mass in our future simulations.
In our present study we have $m_{\pi} L_s\simeq5$ to be compared with
$m_{\pi} L_s\simeq3$ in our previous study. In summary, the pion mass is 
larger but finite volume effects are better under control in the present 
study.

\begin{figure}[t]
\includegraphics[scale=0.4,clip=false]{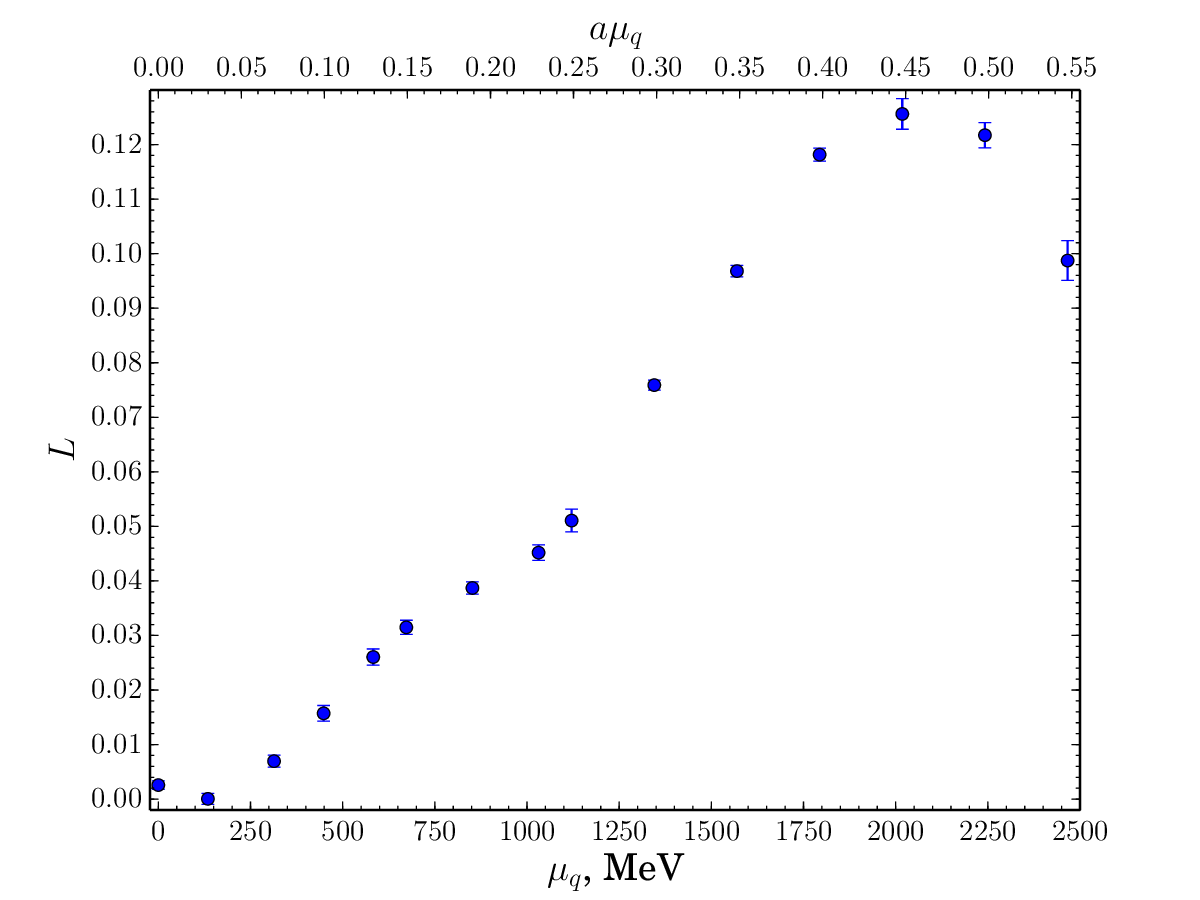}
\caption{The Polyakov loop $\langle L \rangle$ as a function of the quark
chemical potential.
The quark chemical potential is shown in physical and lattice units.}
\label{fig:polyakov_line}
\end{figure}

For the calculation of Wilson loops we have employed one step of HYP 
smearing~\cite{Hasenfratz:2001hp} for temporal links with the smearing 
parameters according to the HYP2 parameter set~\cite{DellaMorte:2005nwx},  
followed by 24 steps of APE smearing~\cite{Albanese:1987ds} for spatial 
links only with the smearing parameter $\alpha_{APE} = 0.25$. The same 
smearing scheme was applied in the paper~\cite{Bonati:2014ksa} for the 
extraction of $V_{Q \bar Q}$ from the Wilson loops. In the case of spatial 
Wilson loops (see below) the smearing scheme was adopted respectively to 
consider one of the spatial directions as a ``temporal direction''. 
For the calculation of the Polyakov loop one step of HYP smearing with the 
same parameters was employed.

We start the presentation of the results of our study with the measurements 
of the average Polyakov loop $\langle L \rangle$. The results are shown in 
Fig.~\ref{fig:polyakov_line}. It is seen from this figure that -- contrary 
to the behaviour of the Polyakov loop at the temperature-driven 
confinement/deconfinement transition where it is a monotonous function of 
temperature -- the dependence of the Polyakov loop on the chemical potential 
is rather complicated. First it rises for chemical potential values
up to $\mu_q \sim 850 \mathrm{~MeV}$ ~($a\mu_q \sim 0.19$). Then 
there is a rapid change of the slope in the region $\mu \in (850,1100)\mathrm{~MeV}$. 
From the chemical potential $\mu_q\sim 1100 \mathrm{~MeV}$ ~($a\mu_q > 0.25$), the Polyakov loop rises 
again reaching a maximum at 
$\mu_q\sim 2000 \mathrm{~MeV}$ ~($a\mu_q \sim 0.45$), before it drops.
It is not quite clear what physical phenomena are hidden behind this
highly nontrivial behaviour of the Polyakov loop.

\begin{figure}[t]
\includegraphics[scale=0.65,clip=false]{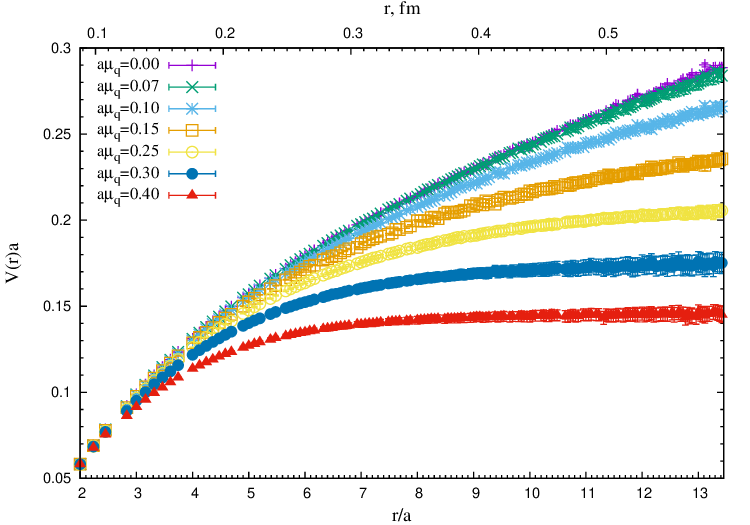}
\caption{The interaction potential $V(r)$ between a static quark-antiquark pair.
Both the potential and the distance between sources are presented in lattice 
units.
}
\label{fig:potential}
\end{figure}

To enquire a possible deconfinement transition in dense matter we measured
the interaction potential between a static quark-antiquark pair through 
the measurement of the Wilson loop. The outcome of this measurement is 
shown in Fig.~\ref{fig:potential}. From this figure it is seen that for 
sufficiently small $\mu_q$ the potential $V(r)$ is a linearly rising function 
of distance, i.e. the system is in the confinement phase. For large values 
of $\mu_q$ the potential $V(r)$ goes to plateau at large distance, i.e. 
the system is in the deconfinement phase. This is in which sense we observe 
a confinement/deconfinement transition in dense matter.

Further let us closer determine the chemical potential characterizing the 
transition from the confinement to the deconfinement phase. To do this we 
find the string tension $\sigma$ as a function of the chemical potential
through a fit of our data by the Cornell potential $V(r)=\alpha/r+\sigma r+c$.
In the fitting procedure we impose the constraint $\sigma \geq 0$.
It turns out that at nonzero chemical potential the value of $\sigma$
depends on the range of distances $r$ covered by the fit. We fit our data in 
the region $r/a \in [5,15]$ and account for additional uncertainty due to 
the variation of the fitting range.  The fit is good for not very large 
values of the chemical potential 
$\mu_q < 1100 \mathrm{~MeV}$ ~($a\mu_q < 0.25$). 
For a chemical potential $\mu_q \geq 1100 \mathrm{~MeV}$ the Cornell potential 
does not describe our data at all.

\begin{figure}[t]
\includegraphics[scale=0.4,clip=false]{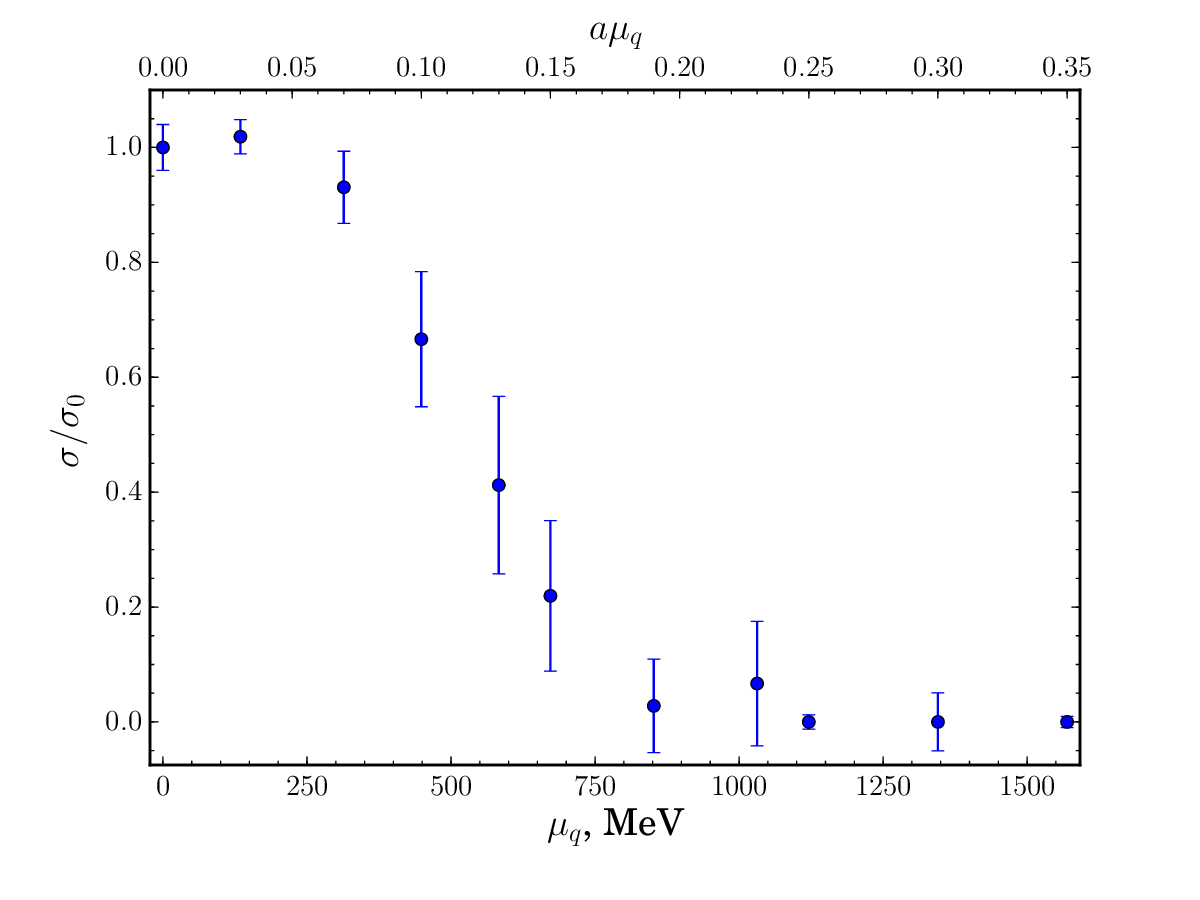}
\caption{The ratio $\sigma/\sigma_0$ as a function of the quark chemical
potential.
The constant $\sigma_0$ is the string tension at $\mu_q=0$.
The quark chemical potential
$\mu_q$ is shown in physical and lattice units.}
\label{fig:sigma}
\end{figure}

In Fig.~\ref{fig:sigma} we plot the ratio $\sigma/\sigma_0$ as a function of 
the chemical potential.
One can see from Fig.~\ref{fig:sigma} that the string tension $\sigma$ 
decreases with increasing chemical potential. Thus we see that the system 
becomes less confined the larger the net quark density is. 
Finally, in the region $\mu_q \geq 850 \mathrm{~MeV}$ ~($a\mu_q \geq 0.19$)
within the uncertainty of the calculation the string tension is zero.
So, we conclude that the deconfinement takes place at 
$\mu_q \geq 850 \mathrm{~MeV}$.

As was mentioned above the Cornell potential describes our data quite well
at sufficiently small chemical potential and it does not describe the data 
at large chemical potential. We believe that this happens since at large 
chemical potential the system under study is in the deconfinement phase 
where the Cornell potential is not applicable.
It is known from $\mu_q =0$ studies that in the deconfinement phase
the static potential of a quark-antiquark pair can be described by the
Debye screened potential $V(r)=(\alpha/r) e^{-m_D r} + c$.
We fitted our data by a Debye--screened Coulomb potential and found that the 
fit is good for values $\mu_q \geq 1100 \mathrm{~MeV}$~($a\mu_q \geq 0.25$).
In this region the screening mass $m_D \neq 0$. In the region 
$\mu_q < 850 \mathrm{~MeV}$ the Debye potential does
not describe our data. Finally, in the region 
$850\mathrm{~MeV}<\mu<1100\mathrm{~MeV}$ the Debye potential fits data quite well but the $m_D$ equals zero within the uncertainty.
Notice that in the region $850\mathrm{~MeV}<\mu<1100\mathrm{~MeV}$ the string
tension $\sigma$ is also zero within the uncertainty of the calculation. 
This implies that our data do not allow us to distinguish confinement phase 
from deconfinement phase in this region. So, one can conclude that the 
confinement/deconfinement transition takes place somewhere in the region 
$\mu_q \in (850,1100) \mathrm{~MeV}$. Below we will take the midpoint of 
this interval $\mu_q\sim1000 \mathrm{~MeV}$ as an estimate of the position 
of the transition.

A more accurate determination of the confinement/deconfinement transition
in dense matter might be obtained through the measurement of the susceptibility
of the Polyakov loop (as well as other susceptibilities).
Unfortunately we are not able to do this since this would require huge 
statistics which is beyond our presently accessible resources.

In order to study how our results are affected by further decreasing 
the temperature of the system we conducted numerical simulations on the
lattice $32^3\times48$ with the same parameters for few values of the 
chemical potential $a\mu_q=0.1, 0.2, 0.3, 0.35$.  We found that the static 
potentials obtained in these simulations are equal to those obtained on the 
lattice $32^4$ within the uncertainty of the calculation. In other words 
the point $a\mu_q=0.1$ is in confinement, the points $a\mu_q=0.3, 0.35$ 
are in deconfinement and the point $a\mu_q=0.2$ is in the transition
region. From this study one can conclude that at low temperatures our 
results concerning confinement/deconfinement transition are not sensitive 
to the temperature.

To get more insight into the confinement/deconfinement transition in dense
matter let us study the dependence of the spatial string tension $\sigma_s$
on the chemical potential. For this purpose we have measured spacelike Wilson 
loops. Taking one spatial direction as a ``time'' direction, one can use the 
relation between the Wilson loop and the static potential and determine a
``spatial potential'' $V_s(r)$. To determine the spatial string tension
we fit $V_s(r)$ by the Cornell parametrization 
$V_s(r)=\alpha_s/r+\sigma_s r+c_s$, too. For all values of the chemical 
potentials under study this form of Cornell potential fits our data well. 
The fitting parameter $\sigma_s$ is the spatial string tension.
Indeed at large distance the spacelike potential $V_s(r)$ rises as $\sigma_s r$,
what implies that the spatial Wilson loop $W_s$ behaves as 
$W_s \sim \exp (-\sigma_s A)$, where $A$ is a area of the surface spanned 
by the Wilson loop.

\begin{figure}[t]
\includegraphics[scale=0.4,clip=false,angle=0]{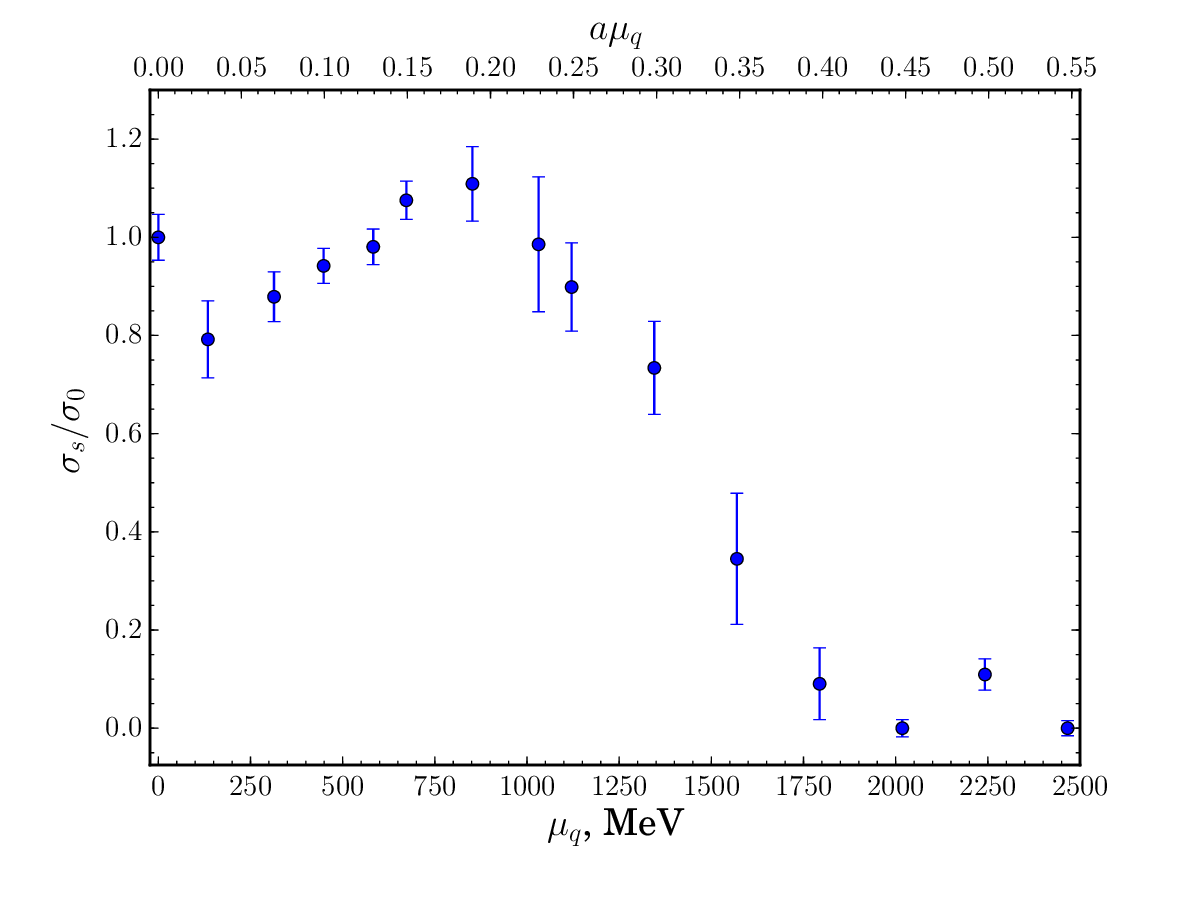}
\caption{The ratio $\sigma_s/\sigma_0$ (of the spacelike string tension)
as a function of the chemical potential.
The constant $\sigma_0$ is the string tension at $\mu_q=0$.
The quark chemical potential $\mu_q$ is shown in physical and lattice units.}
\label{fig:sigmas}
\end{figure}

In Fig.~\ref{fig:sigmas} we plot the ratio $\sigma_s/\sigma_0$ as a function
of the chemical potential. From this Figure one recognizes a monotonous 
decrease of the spatial string tension in the region 
$\mu_q > 1000 \mathrm{~MeV}$. Notice that this decrease starts precisely
in the region where we have observed the conventional (timelike) 
confinement/deconfinement transition in our system. The monotonous decrease 
ends at
$\mu_q \sim 2000 \mathrm{~MeV}$ ~($a\mu_q \sim 0.45$) where the spatial
string tension has become zero within the uncertainty of the calculation.
Thus starting from $\mu_q \sim 2000 \mathrm{~MeV}$ spatial confinement
completely disappears. 
Notice also that in the region $a\mu \in (0, 0.1)$ the spatial string tension 
is smaller than that at zero chemical potential. It is still to be 
clarified if this is a physical effect or statistical fluctuation.

To understand the physical meaning of this result let us recall that the
confinement/deconfinement transition at finite temperature and zero chemical 
potential is connected with the disappearence of the (conventional, timelike)
string tension. On the opposite, above the transition the spatial string 
tension does not vanish. In contrast, it rises with the temperature, and the 
spatial potential in QCD has nonzero string tension $\sigma_s$ at any 
temperature~\cite{Cheng:2008bs}. This means that there are nonperturbative 
effects (in the magnetic sector) in zero density QCD for any temperature.
From this perspective, the confinement/deconfinement transition at finite
density and zero temperature looks quite different. Similarly we observe 
the vanishing of the string tension. Beyond this net quark (baryonic) density 
the spatial string tension starts to decrease. Finally, at the chemical 
potential $\mu_q \sim 2000 \mathrm{~MeV}$ it also vanishes.
Notice that the vanishing of both string tensions indicates the disappearance
of all nonperturbative effects in QCD and, in particular, the disappearance 
of the magnetic screening mass \cite{Son:1998uk}.
At the same time, due to asymptotic freedom, the coupling constant is
already sufficiently small in the region $\mu_q \ge 2000 \mathrm{~MeV}$.
So, one may conjecture that in the region beyond 
$\mu_q \sim 2000 \mathrm{~MeV}$ the quark-gluon plasma is essentially a 
weakly interacting gas of quarks and gluons without magnetic screening 
mass governing the system.

The picture of the confinement/deconfinement transition in dense matter
presented so far in this paper is also supported by our study
~\cite{Bornyakov:2017}
of Abelian monopoles which are revealed by Abelian projection (in the
Maximally Abelian Gauge).
It is known that the percolation properties of the Abelian monopoles
are related to the temperature--driven confinement/deconfinement 
transition. In particular, in the confinement phase there is one 
percolating cluster of monopole currents which disappears at the 
transition to deconfinement. The same behaviour of the monopole 
system is accompanying the density--driven confinement/deconfinement 
transition in dense matter.
In our study we observe always a percolating cluster in the confinement
phase (at small chemical potential). It disappears at 
$\mu_q > 1000 \mathrm{~MeV}$.

It should be added that at high temperature the spatial string tension
is related to the density of monopoles trajectories wrapped around the
(periodic) temporal extent of the lattice. The larger the density of 
wrapped monopoles trajectories the larger is the spatial string tension.
In our study of dense $SU(2)$ QCD we observe that above the 
confinement/deconfinement transition at $\mu_q > 1000 \mathrm{~MeV}$
the density of wrapped monopoles trajectories starts to decrease.
This behaviour is in agreement with the decrease of the spatial string
tension in the same region.

Few words about lattice artifacts are in order. It is known that at
large values of the chemical potential $a\mu_q \sim 1$ a saturation 
effect starts to be seen. The essence of this effect is that all lattice 
sites are filled with fermionic degrees of freedom, and it is not possible 
to put more fermions on the lattice (``Pauli blocking''). It is known that 
the saturation effect is accompanied by the decoupling of the gluons from 
fermions. Thus, effectively due to saturation, our system becomes gluodynamics
which is confined at low temperatures. So, if confinement is seen to persist
at large values of the chemical potential, this is not more than finite lattice spacing artefact. From this consideration it is clear that -- in order to 
successfully observe deconfinement in dense matter -- one should have 
sufficiently small lattice spacing such that the deconfinement is not spoiled 
by this kind of artificial confinement at large values of the chemical 
potential. We believe that for this reason the deconfinement in dense $SU(2)$ 
matter could never be proven before.

Now let us return to our results. 
It may seem from  Fig.~\ref{fig:polyakov_line} that the decrease of the 
Polyakov loop for $\mu>2000$~MeV might be explained by approaching to the 
artificial confinement described above. However, we believe that this is not 
the correct explanation for the following reasons.
First, for $\mu>2000$~MeV (up to $\mu \sim 2500$~MeV) the spatial string 
tension is vanishing. Second, we do not see a respective rise of the timelike
string tension. Moreover, the potential $V(r)$ for $\mu>2000$~MeV is well 
described by Debye screening potential. So, the properties of the system 
in the range $\mu>2000$~MeV are very different from those of plain gluodynamics 
at small temperatures. For this reason we believe that in the region under 
consideration in this work, $\mu<2500$~MeV ($a\mu\leq0.55$) our results are 
not spoiled by eventual saturation effects.

The results of this paper lead us to conclude that the 
confinement/deconfinement transition takes place at 
$\mu_q \sim 1000 \mathrm{~MeV}$. A more detailed study of the position of 
the transition, extending this study to more realistic pion masses, will be 
the goal of our forthcoming study.

Our study of the confinement/deconfinement transition was only possible
in lattice $SU(2)$ QCD. At the end of this paper let us discuss the 
applicability of our results for the case of $SU(3)$ QCD. As was noted above
there are two very important differences between two-color and three-color
QCD. The first one is that the chiral symmetry breaking pattern of the two
theories is different. The second one is that baryons in the $SU(3)$ theory 
are fermions and contain three quarks whereas in the $SU(2)$ theory baryons 
are bosons and contain two quarks.
According to our results the confinement/deconfinement transition takes
place at very large baryon density ($\mu_q \sim 1000 \mathrm{~MeV}$).
Notice that in this region of $\mu_q$ the chiral symmetry is
already restored~\cite{Braguta:2016cpw}. So, we believe that the chiral
symmetry breaking pattern does not play any role in this region. Moreover,
according to our previous results~\cite{Braguta:2016cpw} in the region
of large baryon density the key degrees of freedom are quarks rather
than baryons. Notice also that the ratios of the critical temperature of 
the confinement/deconfinement transitions to the string tension 
$T_c/\sqrt \sigma_0$  in two-color and three-color QCD are close to each other.
These facts allow us to conjecture that if the mechanism of the deconfinement 
in the cold dense matter is the same in SU(2) and SU(3) theories 
the confinement/deconfinement transition in $SU(3)$ theory takes place in 
the region $\mu_q/\sqrt \sigma_0 \sim 2.1$. 
We can also conjecture that the physical scenario of the transition in the $SU(3)$
theory is similar to that in the $SU(2)$ case. In particular, we expect that
the  vanishing of the string  tension is followed by the vanishing of the 
spatial string tension at sufficiently large baryon density also in $SU(3)$ QCD.

In conclusion, in this paper we have studied the confinement/deconfinement
transition aspects in dense lattice $SU(2)$ QCD. The simulations were performed
on a space-time symmetric lattice $32^4$ with rooted staggered fermions at
lattice spacing $a = 0.044$ fm. The small lattice spacing has allowed us to
reach the region of very large baryon density ($\mu_q > 2000 \mathrm{~MeV}$) 
without getting results spoiled by lattice artifacts.
We have measured the Polyakov loop, the interaction potential between static
quark-antiquark pair,
the string tension and the spatial string tension for different values of
the quark chemical potential.

In the region $\mu_q \sim 1000 \mathrm{~MeV}$ we have observed the
confinement/deconfinement transition
which manifests itself in rising of the Polyakov loop and vanishing
of the string tension.
After the onset of deconfinement $\mu_q>1000 \mathrm{~MeV}$ we have 
observed a monotonous decrease of the spatial string tension which ends 
with the vanishing of  this observable in the region 
$\mu_q > 2000 \mathrm{~MeV}$. 
Thus, the confinement/deconfinement transition
at finite density and zero temperature is quite different from that at 
finite temperature and zero density. In addition one may expect that in
very dense matter the quark-gluon plasma becomes a weakly interacting gas
of quarks and gluons without magnetic screening mass in the system, much
different from the quark-gluon plasma at large temperature.

\section*{Acknowledgments}

This work (generation of configurations, computation of the static potential 
and the Polyakov loop) was supported by grant of the Russian Science Foundation
(project number 15-12-20008).
A.Yu.K acknowledges the support from Dynasty foundation. The work was partially
supported by RFBR grant 16-32-00048.
This work has been carried out using computing resources of the federal 
collective usage center Complex for Simulation and Data Processing for 
Mega-science Facilities at NRC ``Kurchatov Institute'',\url{http://ckp.nrcki.ru/}. In addition, we used the supercomputer of the
Institute for Theoretical and Experimental Physics (ITEP).

\bibliographystyle{apsrev4-1}
\bibliography{paper}

\end{document}